\documentclass[%
 aip,
 amsmath,amssymb,
 reprint,%
]{revtex4-1}

\usepackage{graphicx}
\usepackage{dcolumn}
\usepackage{bm}
\usepackage{float}
\usepackage[utf8]{inputenc}
\usepackage[T1]{fontenc}
\usepackage{mathptmx}
\usepackage{etoolbox}

\makeatletter
\def\@email#1#2{%
 \endgroup
 \patchcmd{\titleblock@produce}
  {\frontmatter@RRAPformat}
  {\frontmatter@RRAPformat{\produce@RRAP{*#1\href{mailto:#2}{#2}}}\frontmatter@RRAPformat}
  {}{}
}%
\makeatother
\begin{document}

\preprint{AIP/123-QED}

\title{Influence of grain morphology and orientation on saturation magnetostriction of polycrystalline Terfenol-D}
\author{P. Nieves}
 \email{pablo.nieves.cordones@vsb.cz}

\author{D. Legut}%
\affiliation{IT4Innovations, V\v{S}B - Technical University of Ostrava, 17. listopadu 2172/15, 70800 Ostrava-Poruba, Czech Republic}

\date{\today}

\begin{abstract}
In this work we computationally study the effect of microstructure on saturation magnetostriction of Terfenol-D (Tb$_{0.27}$Dy$_{0.73}$Fe$_{2}$) by means of Finite Element Method. The model is based on the equilibrium magnetoelastic strain tensor at magnetic saturation, and shows that the crystal orientation might play a more significant role on saturation magnetostriction than the morphology of the grains. We also calculate the dependence of saturation magnetostriction on the dispersion angle of the distribution  of grains in the oriented growth  crystal directions $<011>$ and $<111>$, finding that not highly oriented grain distributions reduce saturation magnetostriction significantly. This result evinces the importance of
high-quality control of grain orientation  in the  synthesis of grain-aligned polycrystalline Terfenol-D, and provides a quantitative estimation for the range of acceptable values for the dispersion angle of the distribution  of the oriented grains. 
\end{abstract}

\maketitle

\section{\label{sec:intro}Introduction}

Terfenol-D (Tb$_{0.27}$Dy$_{0.73}$Fe$_2$) is a compound with Laves phase C15 structure type (face centered cubic) that exhibits a  giant magnetostriction along $<111>$ crystallographic direction ($\lambda_{111}=1.6\times10^{-3}$) under moderate magnetic fields ($<2$ kOe) at room temperature \cite{Eng}. Presently, this material is commercially used in many types of sensors and actuators \cite{Dapino}  with different forms like  monolithic rods \cite{Eng}, particle-aligned polymer matrix composites \cite{Anjanappa_1997,altin,McKnight_2002}, and thin films \cite{body}. While single-crystal magnetostrictive alloys are easier to understand and characterize, polycrystals are more suitable for macroscale applications due to lower production cost and faster production rate. The elastic and magnetoelastic properties of polycrystals can be significantly influenced by many factors, such as crystal orientation, orientation degree, microstructure,  volume fraction, grain boundaries, grain morphology and crystal defects. For instance, the low magnetostriction along the  crystallographic direction $<100>$  ($\lambda_{100}=9\times10^{-5}$) of Terfenol-D reduces the saturation magnetostriction up to $\lambda_s\simeq 1\times10^{-3}$ in polycrystalline samples with randomly oriented grains \cite{Dapino}. 

The orientation of the grains in some specific crystallographic directions can significantly modify the elastic and magnetoelastic properties of polycrystalline Terfenol-D \cite{savage1979,Geoffrey}. A typical configuration in monolithic Terfenol-D is to align the crystal axis $<112>$ along the rod axis\cite{Kelogg2007,Dapino,Eng}. Other possible growth directions have also been studied like $<110>$ and $<111>$\cite{Guang-heng,wang2005,JI2002291,Yeou2006,Busbridge}. For instance, Wang et al. investigated Terfenol-D polycrystals with $<110>$ axial alignment finding good magnetostrictive properties  under low magnetic fields \cite{wang2005}. Grain-aligned Terfenol-D in growth direction $<110>$ and $<112>$ can be made with different  directional solidification techniques, such as Bridgman method, Czochralski method, and zone melting method \cite{Deng2006}, while in growth direction $<111>$ Terfenol-D has been successfully prepared using a seeding technique\cite{Guang-heng}. The grain's morphology and orientation of the samples, as well as resulting magnetoelastic properties, are very sensitive to synthesis conditions such as cooling rate, crystal growth velocity, magnetic field, temperature of melt, chemical composition and temperature gradient \cite{Deng2006,Deng200639,Palit2008819,Ji2002309,JI2002291}. Different grain's morphology has been reported in Terfenol-D like plate-shaped\cite{JI2002291}, round-shaped, polygons\cite{Ji2002309} and column-shaped\cite{Deng2006}, but the influence and role of these geometries on saturation magnetostriction are still unclear. In Terfenol-D composites it was found that shape anisotropy induced by the morphology of the grains could be responsible for the grain orientation in a magnetic field\cite{McKnight_2002}. On the other hand, 
 a statistical distribution of the grain orientation around a preferred growth direction is likely to take place in systems with many aligned grains. This property might be described by the dispersion angle of the crystal orientation distribution around a preferred growth direction. For instance, highly oriented grains in $<110>$ with dispersion angle of only $3^\circ$ has been achieved using a temperature gradient directional solidification technique\cite{JI2002291}. The influence of this dispersion angle on saturation magnetostriction has not been quantified in much detail yet. This correlation could provide some hints about the range of acceptable values for the quality of the grain orientation in the synthesis of grain-aligned polycrystalline Terfenol-D. We point out that small deviations from a perfect alignment might induce significant changes on saturation magnetostriction due to the large difference between the magnetostriction in the crystallographic directions $<111>$ and $<100>$ ($\lambda_{111}\gg\lambda_{100}$).  Here, we try to estimate these effects  on saturation magnetostriction computationally.

Terfenol-D exhibits a complex nonlinear stress-strain relationship\cite{kellog}, where its Young’s modulus depends  on the stress and magnetic field strongly. In general, one needs advanced nonlinear constitutive models to describe all magnetostrictive features of this material\cite{zheng2005,Sudersan2020}. In this work, we restrict our study only to the magnetic saturated state of Terfenol-D, which allows us to greatly simplify the Finite Element Method (FEM) model through the use of the  theoretical equilibrium magnetoelastic strain tensor at saturation\cite{CLARK1980531}. 

\begin{figure*}
\centering
\includegraphics[width=1.5\columnwidth ,angle=0]{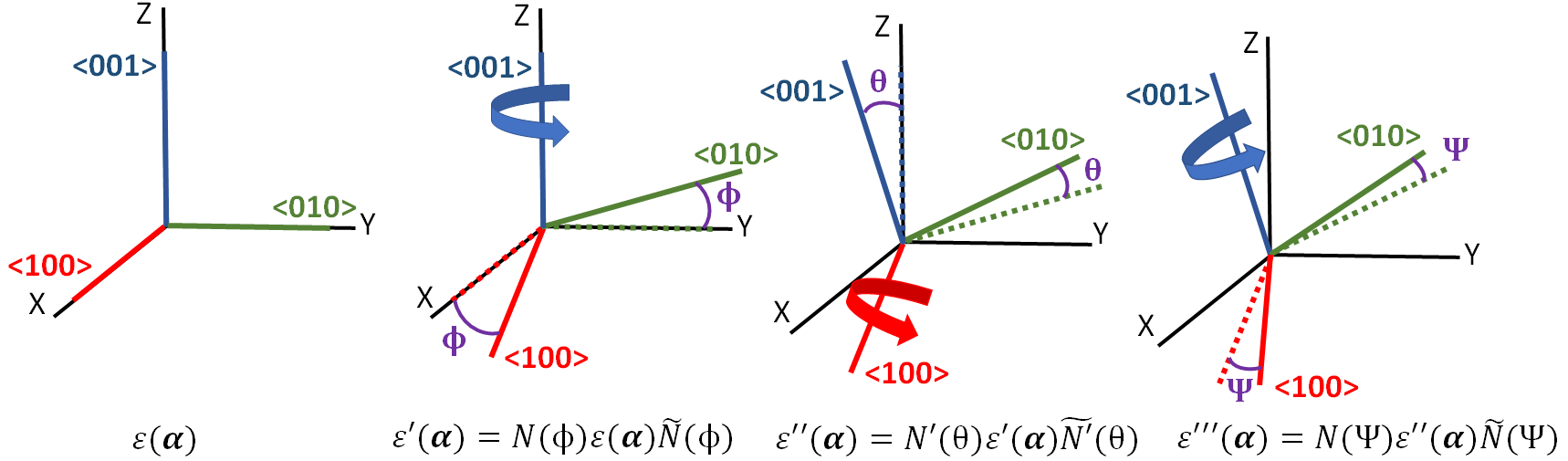}
\caption{Transformation of the magnetoelastic strain tensor ($\epsilon$) when an active rotation of the crystal is performed with Euler angles $(\phi,\theta,\psi)$ in Bunge convention.}
\label{fig:rot_C_eps}
\end{figure*}

\section{Methodology}
\label{section:method}

\subsection{Theoretical equations}
\label{section:theory}

To compute saturation magnetostriction, we use the equilibrium magnetoelastic strain tensor that can be analytically derived from the minimization of both the elastic ($E_{el}$) and magnetoelastic ($E_{me}$) energies with respect to strain $\epsilon$ ~\cite{CLARK1980531,Cullen}
\begin{equation}
    \frac{\partial(E_{el}+E_{me})}{\partial\epsilon_{ij}}=0. \quad\quad i,j=x,y,z
    \label{eq:dE}
\end{equation}

For a cubic single crystal (point groups $432$, $\bar{4}3m$, $m\bar{3}m$) equilibrium magnetoelastic strain tensor reads~\cite{CLARK1980531,Cullen}
\begin{widetext}
\begin{equation}
\begin{aligned}
 \epsilon (\boldsymbol{\alpha}) = 
\begin{pmatrix}
\lambda^\alpha +\frac{3}{2}\lambda_{001}\left(\alpha_{100}^2-\frac{1}{3}\right) & \frac{3}{2}\lambda_{111}\alpha_{100}\alpha_{010} & \frac{3}{2}\lambda_{111}\alpha_{100}\alpha_{001} \\
\frac{3}{2}\lambda_{111}\alpha_{100}\alpha_{010} & \lambda^\alpha +\frac{3}{2}\lambda_{001}\left(\alpha_{010}^2-\frac{1}{3}\right) & \frac{3}{2}\lambda_{111}\alpha_{010}\alpha_{001} \\
\frac{3}{2}\lambda_{111}\alpha_{100}\alpha_{001}  & \frac{3}{2}\lambda_{111}\alpha_{010}\alpha_{001}  & \lambda^\alpha +\frac{3}{2}\lambda_{001}\left(\alpha_{001}^2-\frac{1}{3}\right)\\
\end{pmatrix},
\label{eq:magelas_strain}
\end{aligned}
\end{equation}
\end{widetext}
where $\lambda^\alpha$ describes the
isotropic magnetostriction, while $\lambda_{001}$ and $\lambda_{111}$ are the anisotropic magnetostrictive coefficients that give the fractional length
change along the $<001>$ and $<111>$ crystallographic directions when a demagnetized material is magnetized in these directions, respectively. The vector  $\boldsymbol{\alpha}=(\alpha_{100}, \alpha_{010},\alpha_{001})$ gives the reduced magnetization (saturated magnetic state) with respect to the crystal axes,  that is, the quantities $\alpha_{100}$, $\alpha_{010}$ and $\alpha_{001}$ denote the components of the reduced magnetization along the crystallographic directions $<001>$, $<010>$ and $<001>$, respectively.  

\begin{figure}[h!]
\centering
\includegraphics[width=\columnwidth ,angle=0]{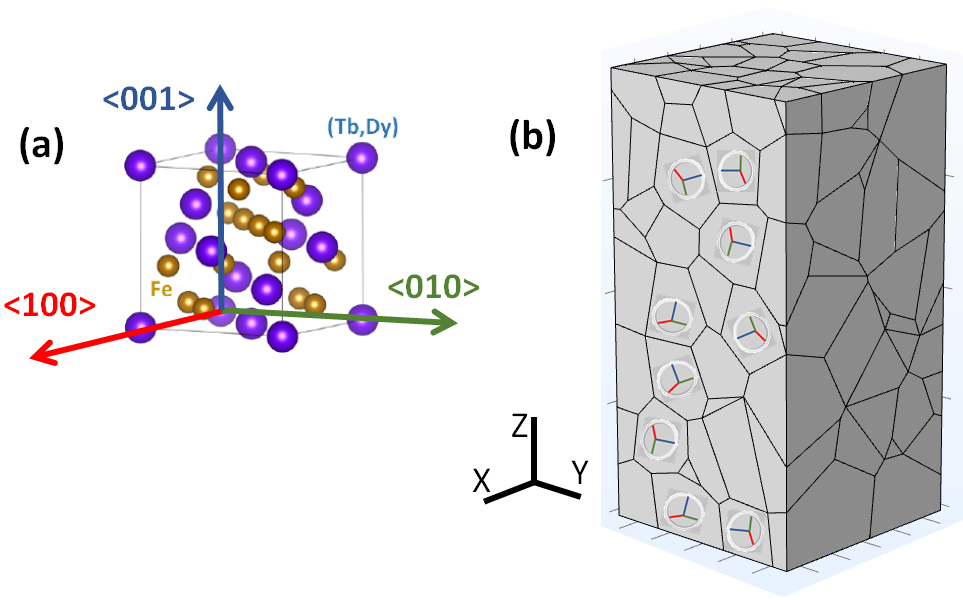}
\caption{(a) Unit cell of Terfenol-D with Laves phase C15 structure. (b) Polycrystalline model of Terfenol-D where the crystallographic axes of some grains are depicted.}
\label{fig:crystal_poly}
\end{figure}

\begin{figure}[h!]
\centering
\includegraphics[width=\columnwidth ,angle=0]{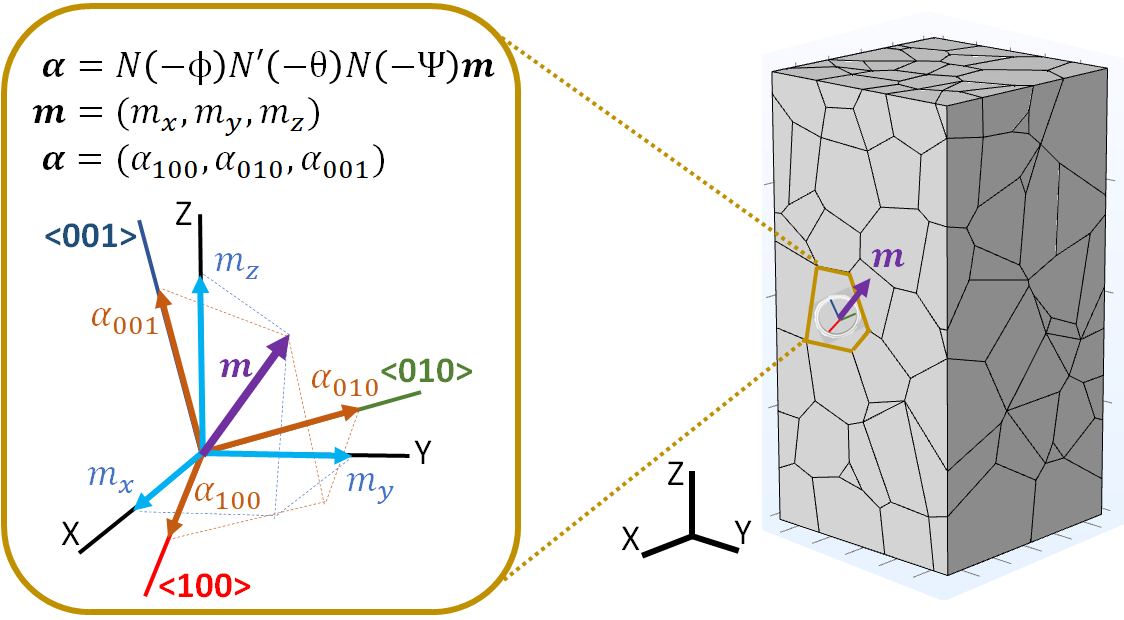}
\caption{Diagram showing the components of the reduced magnetization vector ($\boldsymbol{m}$) for a particular grain inside the polycrystalline material. The components of the reduced magnetization are shown in the global polycrystalline material's coordinates XYZ ($m_x,m_y,m_z$) and rotated crystallographic axes ($\alpha_{100},\alpha_{010},\alpha_{001}$). The grain is actively rotated with Euler angles $(\phi,\theta,\psi)$ in Bunge convention, as depicted in Fig.\ref{fig:rot_C_eps}.}
\label{fig:rot_mag}
\end{figure}

The form of the magnetoelastic strain tensor in Eq.\ref{eq:magelas_strain} corresponds to a cubic crystal with crystallographic axes $<100>$, $<010>$ and $<001>$ along the Cartesian $X$, $Y$ and $Z$ axes, respectively. If we perform an active rotation of the crystal with Euler angles $(\phi,\theta,\psi)$ in Bunge convention $Z$-$X'$-$Z''$, then the magnetoelastic strain tensor is transformed as\cite{Auld}
\begin{equation}
\begin{aligned}
\epsilon'''(\boldsymbol{\alpha})=N(\psi)N'(\theta)N(\phi)\epsilon(\boldsymbol{\alpha})\tilde{N}(\phi)\tilde{N}'(\theta)\tilde{N}(\psi),
\label{eq:epsilon_tensor_cub_rot}
\end{aligned}
\end{equation}
where
\begin{equation}
\begin{aligned}
N(\phi) & =\begin{pmatrix}
cos(\phi) & -sin(\phi) & 0  \\
sin(\phi) & cos(\phi)  &  0  \\
0 & 0 & 1 \\
\end{pmatrix},\\
N'(\theta) & =\begin{pmatrix}
1 & 0 & 0 \\
0 & cos(\theta)  & -sin(\theta)    \\
0 & sin(\theta) & cos(\theta)  \\
\end{pmatrix}.
\label{eq:m_matrix}
\end{aligned}
\end{equation}
The matrices $\tilde{N}$ and $\tilde{N}'$ are the transpose of matrices $N$ and $N'$, respectively. In Fig. \ref{fig:rot_C_eps} we illustrate this transformation, while Fig.\ref{fig:crystal_poly} shows how the crystallographic axes of the unit cell of Terfenol-D is related to the coordinates (XYZ) of the polycrystalline model.  If in a rotated grain inside the polycrystalline model we set the reduced magnetization $\boldsymbol{m}=(m_x,m_y,m_z)$ with respect to the Cartesian axes X, Y and Z, then the reduced magnetization with respect to the crystallographic axes of this grain is given by 
\begin{equation}
\begin{aligned}
\boldsymbol{\alpha}=N(-\phi)N'(-\theta)N(-\psi)\boldsymbol{m}.
\label{eq:mag_rot}
\end{aligned}
\end{equation}
This conversion formula is useful for the evaluation of magnetoelastic strain tensor of each rotated grain in a polycrystalline materials since the magnetization of all grains is saturated to a particular direction with respect to the Cartesian coordinates X, Y and Z of the  whole  material, see Fig.\ref{fig:rot_mag}. 

Once the equilibrium strain tensor of each rotated grain with saturation magnetization in the direction $\boldsymbol{m}$ is computed through Eq. \ref{eq:epsilon_tensor_cub_rot} combined with Eq.\ref{eq:mag_rot}, we can calculate the final deformation that takes place from the demagnetized state to the magnetic saturated state. The deformation is given by the displacement vector $\boldsymbol{u}(\boldsymbol{r})=\boldsymbol{r'}-\boldsymbol{r}$ that gives the displacement from a point at the initial position $\boldsymbol{r}$ (at the demagnetized state) to its final position $\boldsymbol{r'}$ (at the magnetic saturated state). For small deformations, the displacement vector is calculated by solving~\cite{Landau}
\begin{equation}
\begin{aligned}
   \epsilon_{ij}'''(\boldsymbol{\alpha})=\frac{1}{2}\left(\frac{\partial u_{i}}{\partial r_{j}}+\frac{\partial u_j}{\partial r_i}\right),\quad\quad i,j=x,y,z
  \label{eq:disp_vec}
\end{aligned}
\end{equation}
where $\epsilon_{ij}'''$ is the transformed strain tensor (Eq. \ref{eq:epsilon_tensor_cub_rot}) of each rotated grain.

\begin{figure}[h!]
\centering
\includegraphics[width=0.7\columnwidth ,angle=0]{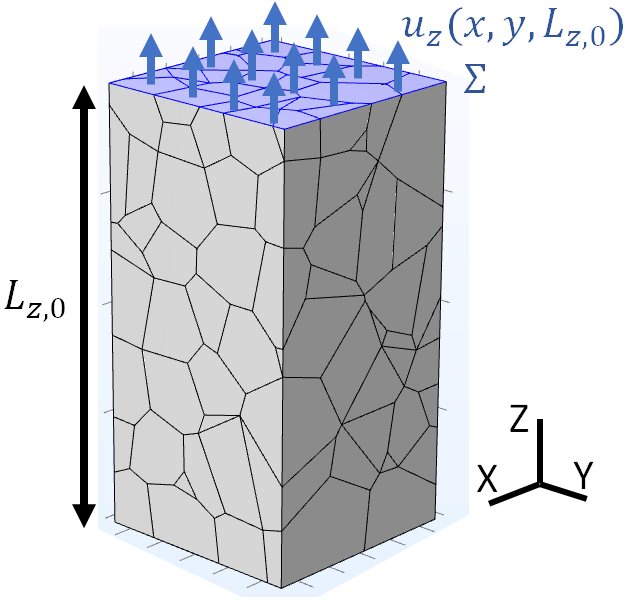}
\caption{Scheme showing the calculation of the change in length along the z-axis for the polycrystalline model of Terfenol-D. }
\label{fig:cal_def}
\end{figure}

\subsection{Numerical methods}
\label{section:num_methods}

In practice, we first compute $\boldsymbol{\alpha}$ for each grain via Eq.\ref{eq:mag_rot} by using the given Euler angles of each grain and the selected direction to saturate the magnetization of all grains $\boldsymbol{m}$. Next, we use the given Euler angles and calculated vector $\boldsymbol{\alpha}$ of each grain to compute the equilibrium strain tensor $\epsilon'''$ via Eq. \ref{eq:epsilon_tensor_cub_rot}. Lastly, the deformation is obtained by solving Eq.\ref{eq:disp_vec}. This procedure is done by means of FEM techniques as implemented in the software COMSOL\cite{comsol} interfaced with MATLAB\cite{MATLAB:2020}. In the simulations we used a tetrahedral mesh and the direct method PARDISO to solve the linear systems of equations  with default settings of COMSOL (version 5.6). The generation of the polycrystalline models and nearly uniform sampling of crystal orientations (Euler angles, active rotation in Bunge convention) are performed with software Neper \cite{neper,Quey:ks5599}. Once the deformation of the system is calculated, the change in length $\Delta L_z$ in the z-axis direction is calculated as the spatial average of the z-component of the displacement vector ($u_z$) on the top of boundary $\Sigma$
\begin{equation}
    <\Delta L_z> = < L_z - L_{z,0} >  = \frac{\int_{\Sigma}u_z(x,y,L_{z,0})dxdy}{\int_{\Sigma}dxdy},
\label{eq:delta_L}     
\end{equation}
where $L_{0,z}$ is the initial length in the z-axis direction in the demagnetized state (before deformation takes place). A fixed constraint is imposed at the bottom boundary $u_z(x,y,0)=0$ for reference purposes, see Fig.\ref{fig:cal_def}. Eq.\ref{eq:delta_L} is numerically computed with the analysis tools implemented in COMSOL \cite{comsol}. Next, magnetostriction is calculated as the relative length change $\lambda=<\Delta L_z>/L_{z,0}$. For the magnetostrictive coefficients of Terfenol-D we use its room temperature values  $\lambda_{001}=90\times10^{-6}$,  $\lambda_{111}=1600\times10^{-6}$ and $\lambda^{\alpha}=0$ \cite{Dapino,ARNAUDAS2002389}.

\begin{figure*}
\centering
\includegraphics[width=1.7\columnwidth ,angle=0]{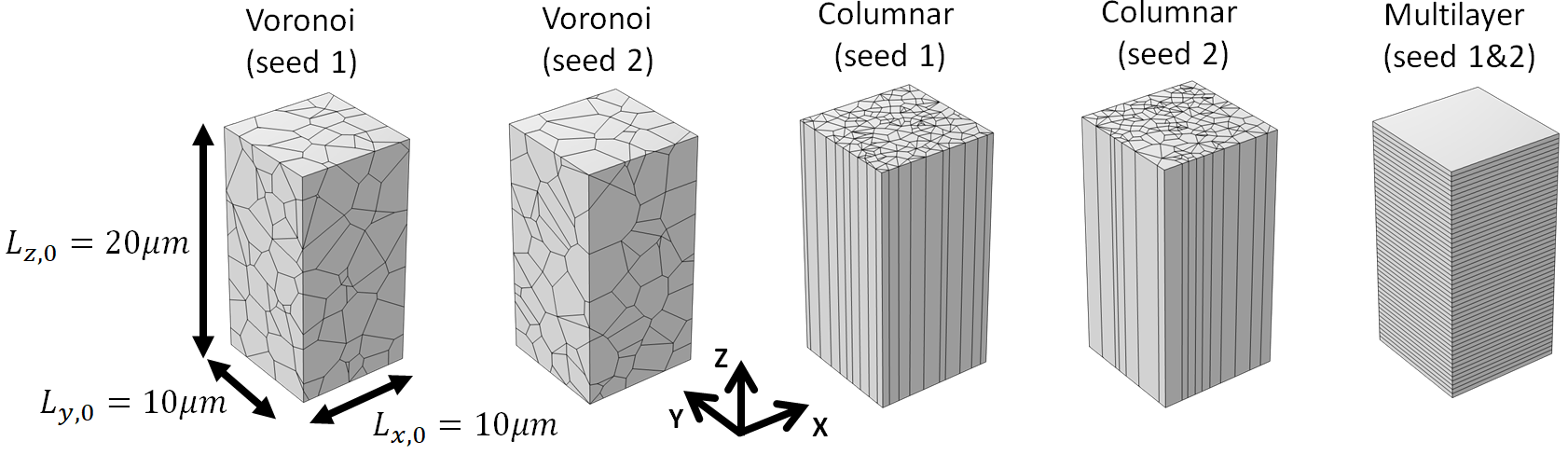}
\caption{Polycrystalline models considered in this work to study the influence of grain morphology on saturated magnetostriction. }
\label{fig:micro}
\end{figure*}

\begin{figure}
\centering
\includegraphics[width=\columnwidth ,angle=0]{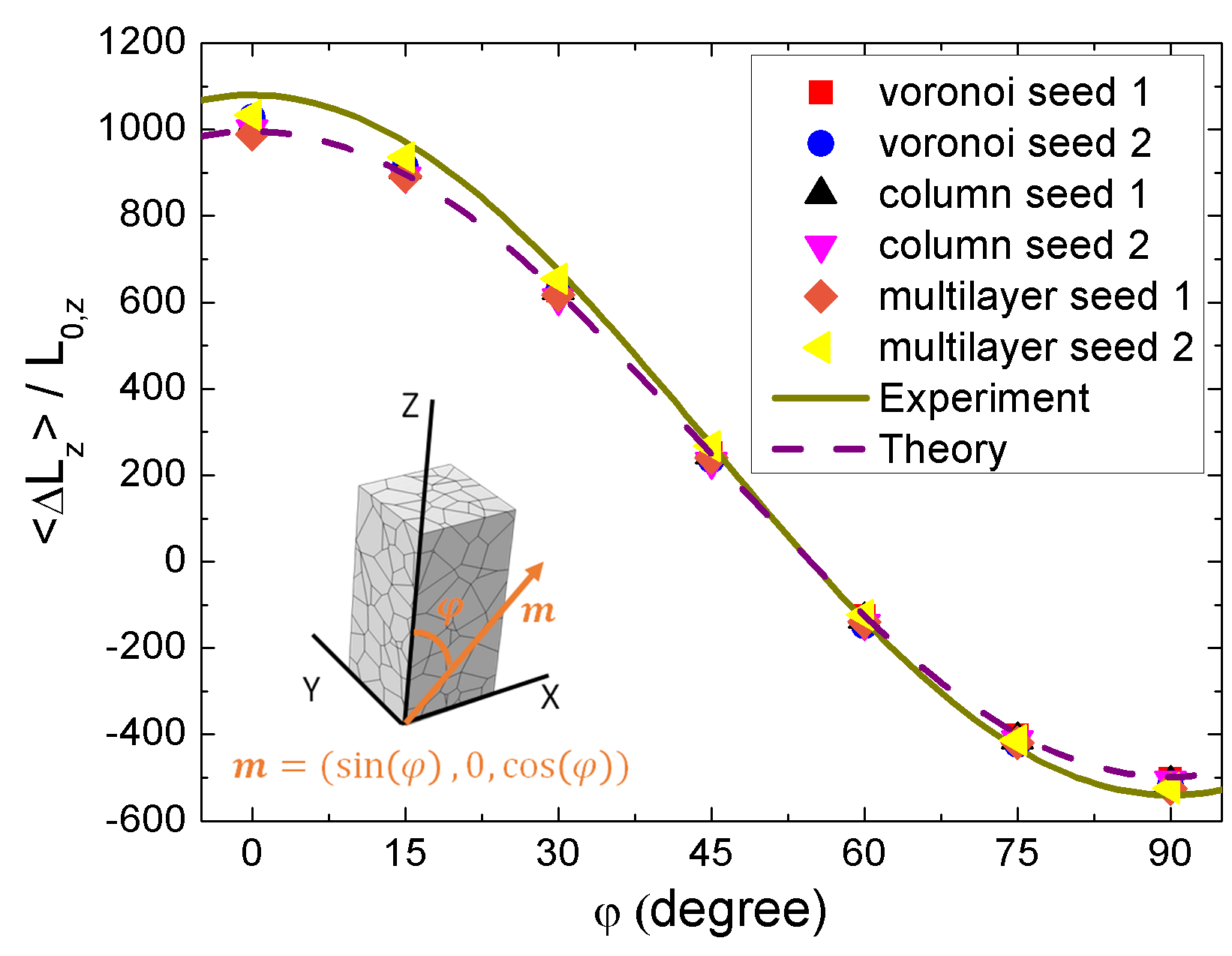}
\caption{Relative length change along z-axis direction versus the angle $\varphi$ between the magnetization direction of all grains $\boldsymbol{m}=(sin\varphi,0,cos\varphi)$ and z-axis direction $\boldsymbol{\beta}'=(0,0,1)$. Symbols represent FEM simulations for different types of grain morphology and uniformly-distributed crystal orientations in 3D space. Solid and dash lines stand for Eq.\ref{eq:delta_l_cub_poly} using the experimental and theoretical value for $\lambda_S$, respectively.}
\label{fig:lmb_poly}
\end{figure}

\begin{figure*}
\centering
\includegraphics[width=1.9\columnwidth ,angle=0]{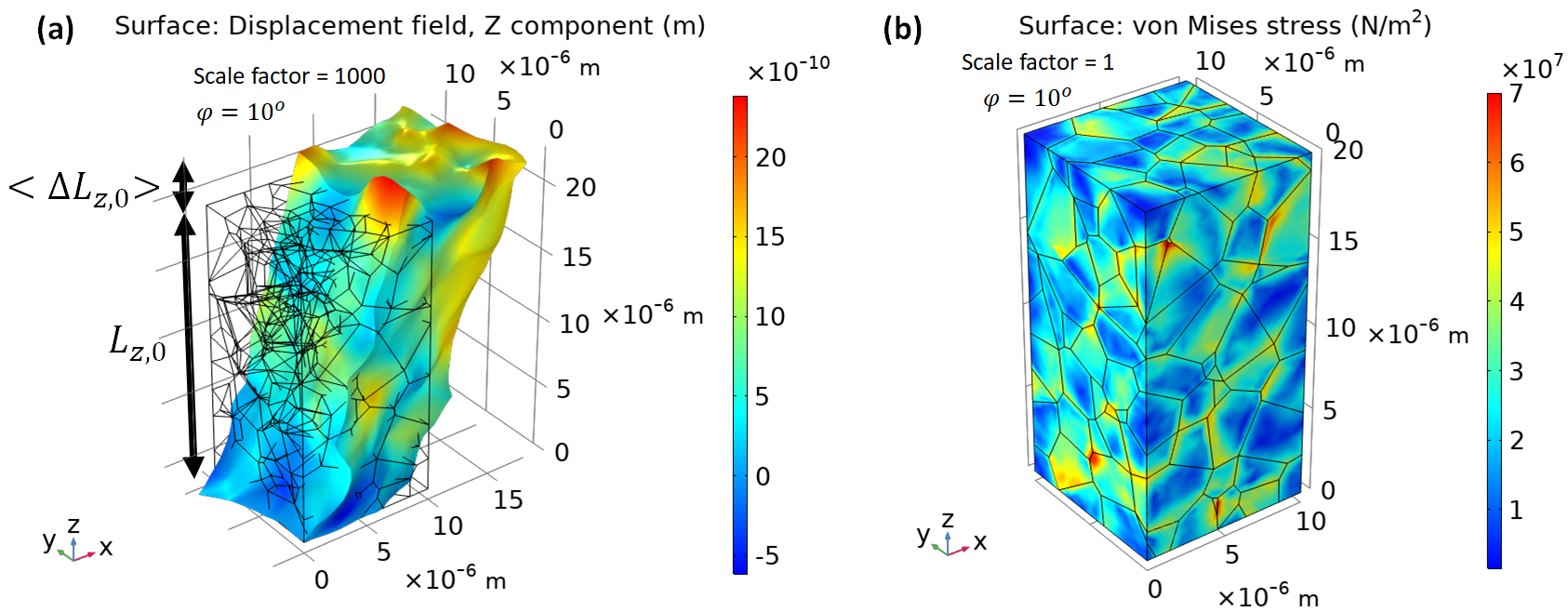}
\caption{Calculated magnetostrictive response for the  voronoi microstructure (seed 1) with uniformly-distributed crystal orientations in 3D space when magnetization of all grains is saturated along $\boldsymbol{m}=(sin\varphi,0,cos\varphi)$ with $\varphi=10^\circ$. Surface color shows (a) the z-component of the displacement vector $u_z$ and (b) von Mises stress. A scale factor $1000$ is applied to (a) in order to facilitate the visualization of the deformation.}
\label{fig:deform_lmb}
\end{figure*}

\begin{table}[]
\centering
\caption{Isotropic magnetostrictive coefficient ($\lambda_S$) calculated with FEM simulations for different types of microstructure with uniformly-distributed crystal orientations in 3D space. The theoreti\-cal and experimental values are also shown for comparison. The theoretical value for $\lambda_S$ is obtained via Eq.\ref{eq:lmb_s}.}
\label{tab:lmb_s}
\begin{ruledtabular}
\begin{tabular}{cc}
Approach &
  \begin{tabular}[c]{@{}c@{}}$\lambda_{S}$ \\ ($\times 10^{-6}$)\end{tabular} \\ \hline
  FEM voronoi (seed 1) & 1005  \\
  FEM voronoi (seed 2) & 1022  \\
  FEM columnar (seed 1) & 1011 \\
  FEM columnar (seed 2) & 997   \\
  FEM multilayer (seed 1) & 998  \\
  FEM multilayer (seed 2) & 1040 \\
  Theory  & 996 \\
  Experiment & 1080$^a$  \\
\end{tabular}
\end{ruledtabular}
\footnotesize $^a$Ref.\cite{Dapino}
\end{table}

\section{Results}
\label{section:results}

\subsection{Influence of grain morphology}
\label{section:morpho}

To study the influence of grain morphology on elastic and magnetoelastic properties, we consider a material made of Terfenol-D with a rectangular geometry ($L_{x,0}=L_{y,0}=10\mu$m and $L_{z,0}=20\mu$m) and 3 types of microstructure: (i) voronoi, (ii) columnar and (iii) multilayer, see Fig.\ref{fig:micro}. For the voronoi and columnar microstructures we include 200 grains with uniformly-distributed orientations in 3D space \cite{Quey:ks5599}, and we study two cases for each microstructure using a different seed in their generation with software Neper \cite{neper}. For the multilayer microstructure we include 50 layers with the same width ($0.4\mu$m), and we study two cases using the same morphology but with different uniformly-distributed crystal orientations of the layers. Following the methodology explained in Section \ref{section:num_methods}, we computed the relative length change along z-axis direction versus the angle $\varphi$ between the saturated magnetization direction of all grains $\boldsymbol{m}=(sin\varphi,0,cos\varphi)$ and z-axis direction, see Fig.\ref{fig:lmb_poly}. We find that the FEM results for all models follow the relation \cite{Birss,Dapino}
\begin{equation}
     \frac{\Delta l}{l_0}\Bigg\vert_{\boldsymbol{\beta}'}^{\boldsymbol{m}} = \frac{3}{2}\lambda_{S}\left[(\boldsymbol{m}\cdot\boldsymbol{\beta}')^2-\frac{1}{3}\right],
    \label{eq:delta_l_cub_poly}
\end{equation}
where $\lambda_{S}$ is the isotropic magnetostrictive coefficient of the polycrystalline material, and the vector $\boldsymbol{\beta}'=(\beta_x',\beta_y',\beta_z')$ is a unit vector that describes the measuring length direction in the global Cartesian coordinates XYZ of the polycrystalline material. For instance, since here we are computing the relative length change along z-axis direction, then we have $\boldsymbol{\beta}'=(0,0,1)$. In the theory of magnetostriction for polycrystalline materials, a widely used approximation is to assume that the stress distribution is uniform through the material. In this case the isotropic magnetostrictive coefficient reads \cite{Akulov,Lee_1955,Cullen,Birss,Dapino}
\begin{equation}
\lambda_S=\frac{2}{5}\lambda_{001}+\frac{3}{5}\lambda_{111}.
 \label{eq:lmb_s}
\end{equation}
This result is analogous to the Reuss approximation used in the elastic theory of polycrystals to obtain a lower bound of Young's modulus \cite{Cullen,Reuss,Hill_1952,AELAS}. In Table \ref{tab:lmb_s} we present the results for $\lambda_S$ calculated with FEM simulations for different types of microstructure with uniformly-distributed crystal orientations in 3D space. These values are obtained by fitting the data shown in Fig.\ref{fig:lmb_poly} to Eq.\ref{eq:delta_l_cub_poly}. The theoretical and experimental values are also shown for comparison in Table \ref{tab:lmb_s}. The theoretical value is obtained via Eq.\ref{eq:lmb_s} with the anisotropic magnetostrictive coefficients used in the model ($\lambda_{001}=90\times10^{-6}$ and  $\lambda_{111}=1600\times10^{-6}$). We see that Eq.\ref{eq:lmb_s} gives a reasonable lower bound for $\lambda_S$ ($\lambda_S=996\times10^{-6}$) that is close to the experimental value\cite{Dapino} $\lambda_S=1080\times10^{-6}$. We observe that all simulated microstructures lead to very similar values for $\lambda_S$. The observed small difference in $\lambda_S$ could be more related to the different random orientation distribution of each model rather than the morphology. This fact is clearly seen for the multilayer models since the same grain morphology is used in the two simulated cases. A more detailed view of the magnetostrictive deformation for the voronoi microstructure (seed 1) is depicted in the Fig.\ref{fig:deform_lmb}. In this figure, we also plotted the von Mises stress that can be used to analyze possible failures of the material under mechanical stress. We see that the highest values of von Mises stress ($\sim 0.07$ GPa) are found close to some specific grain boundaries.

\begin{figure*}
\centering
\includegraphics[width=1.8\columnwidth ,angle=0]{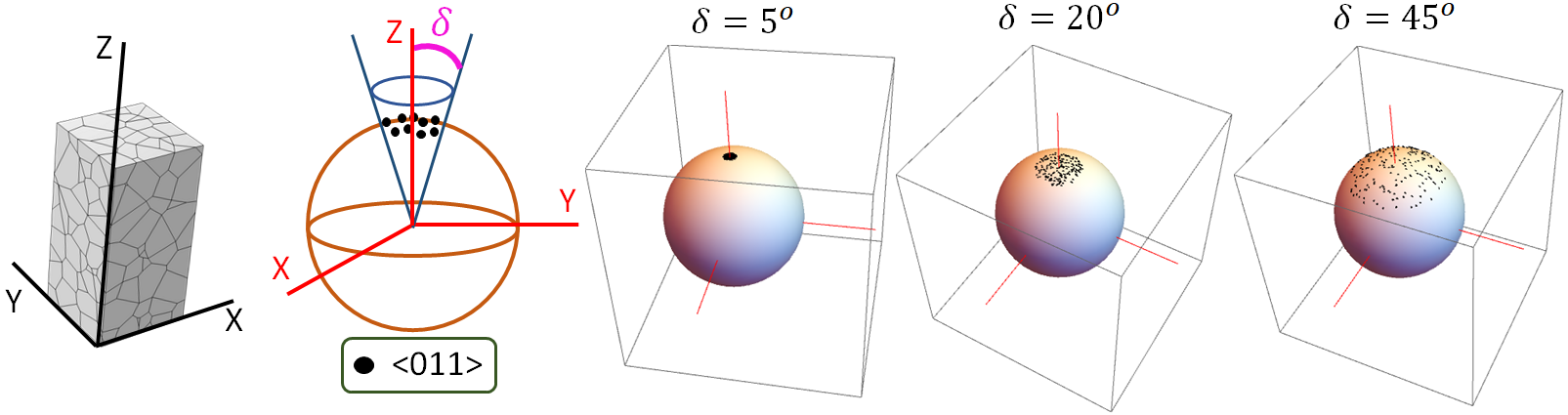}
\caption{Uniformly distribution of crystallographic axis $<011>$ along the z-axis direction with dispersion angle $\delta=5^\circ,20^\circ,45^\circ$.  }
\label{fig:disp_011}
\end{figure*}

\begin{figure}
\centering
\includegraphics[width=\columnwidth ,angle=0]{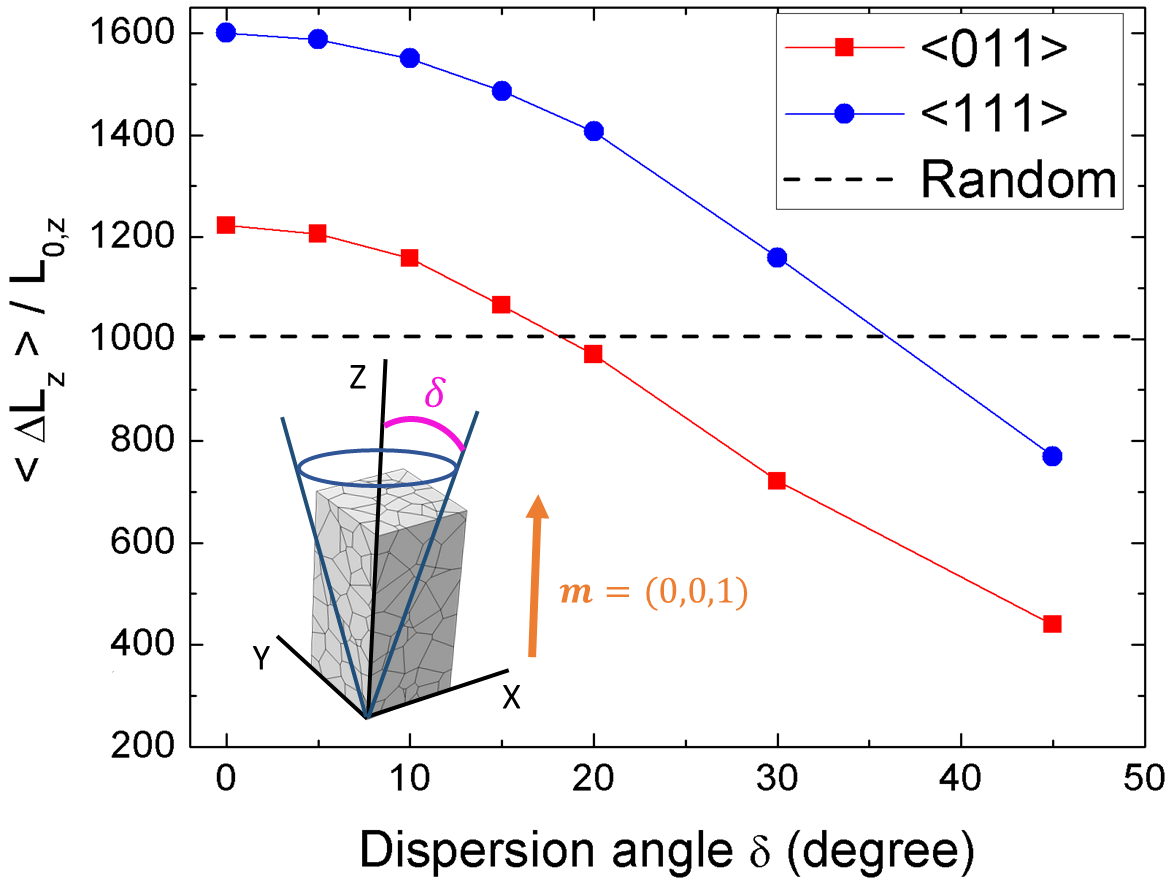}
\caption{Computed relative length change along the z-axis as a function of the dispersion angle $\delta$ for the FEM models with uniformly distribution of crystallographic axes $<011>$ and $<111>$ along the z-axis direction. The magnetization of all grains is saturated along the z-axis $\boldsymbol{m}=(0,0,1)$. Horizontal dash line stands for the result given by the FEM model using a  uniformly distribution of the crystal orientations in all 3D space.  }
\label{fig:Y_lmb_disp}
\end{figure}

\subsection{Influence of grain orientation}
\label{section:ori}

Elastic and magnetoelastic properties of polycrystalline materials can be tuned by orienting the growth of all grains along a specific crystallographic direction \cite{savage1979,Yeou2006,wang2005,JI2002291}. In this section, we computationally explore how this effect influences the saturated magnetostriction for Terfenol-D. For this analy\-sis we consider the same voronoi microstructure (seed 1) than in Section \ref{section:morpho}, but now the crystallographic direction $<011>$ or $<111>$ of all grains will be uniformly oriented to the z-axis with a dispersion angle $\delta$. The magnetization of all grains is saturated along the z-axis $\boldsymbol{m}=(0,0,1)$. The uniformly distribution of the crystallographic direction $<011>$ along the z-axis with a dispersion angle $\delta$ is achieved by setting the Euler angles to rotate each grain as \cite{point_sphere}
\begin{equation}
\begin{aligned}
\vert arccos(2v-1)\vert & < \delta,\\
\phi & = 0,\\
\theta & = \frac{\pi}{4}-\vert arccos(2v-1)\vert,\\
\psi & = 2\pi u,
\end{aligned}
 \label{eq:rot_011}  
\end{equation}
where $v$ and $u$ are random real numbers between $0$ and $1$. In Fig.\ref{fig:disp_011} we show the generated uniformly distribution of crystallographic axis $<011>$ along the z-axis direction with dispersion angle $\delta=5^\circ,20^\circ,45^\circ$. Similarly, the uniformly distribution of the crystallographic direction $<111>$ along the z-axis with a dispersion angle $\delta$ is achieved by setting the Euler angles to rotate each grain as \cite{point_sphere}
\begin{equation}
\begin{aligned}
\vert arccos(2v-1)\vert & < \delta,\\
\phi & = \frac{\pi}{4},\\
\theta & = \pi-arctan\left(\frac{1}{\sqrt{2}}\right)-\vert arccos(2v-1)\vert,\\
\psi & = 2\pi u.
\end{aligned}
 \label{eq:rot_111}  
\end{equation}
In Fig. \ref{fig:Y_lmb_disp} we present the relative length change along the z-axis as a function of the dispersion angle $\delta$ for the FEM models with uniformly distribution of crystallographic axes $<011>$ and $<111>$ along the z-axis direction. We observe that the dispersion angle has a very strong effect on the saturation magnetostriction of Terfenol-D which is favored by the fact $\lambda_{111}>>\lambda_{100}$. As a result, a broad range of values of saturation magnetostriction can be achieved by tuning the dispersion angle. For instance, we see that  orientation distributions in the crystallographic axis $<111>$ with dispersion angle $\delta$ lower than $35^\circ$ exhibit larger magnetostriction than an uniformly distribution of the crystal orientations in the all 3D space. For this case, we verify that in the limit $\delta\xrightarrow{}0$ the relative length change leads to $\lambda_{111}=1600\times10^{-6}$, as it should be.  Interestingly, for  small dispersion angles up to $\delta \approx 10^\circ$ the saturation magnetostriction changes less than 5\% with respect to the perfectly aligned case ($\delta = 0^\circ$). Hence, $\delta \approx 10^\circ$ could be considered as a possible acceptable upper limit for the dispersion angle to achieve high saturation magnetostriction in grain-aligned polycrystalline Terfenol-D.

\section{Conclusions}
\label{section:conclusions}

In summary, our computational analysis of the effect of microstructure on saturation magnetostriction for Terfenol-D shows that crystal orientation distribution could play a more important role than grain morphology. We also quantified the effect of the dispersion angle of the grains on saturation magnetostriction, finding that it influences this property strongly.  We estimated that small dispersion angles up to $10^\circ$ might be sufficient to achieve high saturation magnetostriction in the oriented growth crystal directions $<011>$ and $<111>$ of polycrystalline Terfenol-D. This result shows the need for a precise control of grain orientation distribution in the synthesis of grain-aligned polycrystalline Terfenol-D.

\section*{Acknowledgement}

This work was supported by the ERDF in the IT4Innovations national supercomputing center - path to exascale project (CZ.02.1.01/0.0/0.0/16-013/0001791) within the OPRDE and by The Ministry of Education, Youth and Sports from  the Large Infrastructures for Research, Experimental Development, and Innovations project “e-INFRA CZ (ID:90140)". Also the Donau project No. 8X20050 and support from the H2020-FETOPEN No.~863155 s-NEBULA project is acknowledged.

\section*{Data availability}

The data that support the findings of this study are available from the corresponding author upon reasonable request.

\bibliographystyle{elsarticle-num}
\bibliography{mybibfile}

\end{document}